\documentclass[9pt]{article}

\usepackage{graphicx} 
\usepackage{amsmath}
\usepackage{amssymb}
\usepackage{hyperref}
\usepackage{comment}
\usepackage{float}
\usepackage{caption}
\usepackage{bbold}
\usepackage{amsthm}
\usepackage{booktabs}
\usepackage{authblk}
\usepackage{multirow}
\usepackage{xcolor}
\usepackage{cases}
\usepackage[numbers]{natbib}

\graphicspath{{Images/}}

\usepackage{geometry}
\title{Classical JAK2V617F+ Myeloproliferative Neoplasms emergence and development based on real life incidence and mathematical modeling}
\author[1]{Ana Fern\'andez Baranda}
\author[1]{Vincent Bansaye}
\author[2]{Evelyne Lauret}
\author[3]{Morgane Mounier}
\author[4]{Val\'erie Ugo}
\author[*, 1]{Sylvie M\'el\'eard}
\author[*, 5]{St\'ephane Giraudier}

\affil[*]{\raggedright {\small These two authors are co-last authors.}}
\affil[1]{{\small CMAP, Ecole Polytechnique, CNRS, Institut Polytechnique de Paris, Inria, route de Saclay 91128 Palaiseau, France}}
\affil[2]{{\small Universit\'e de Paris, Institut Cochin, INSERM U1016, CNRS UMR8104, F-75014 PARIS, France }}
\affil[3]{{\small Malignant Hemopathies Registry of Cote d'Or, INSERM U1231, CHU Dijon Bourgogne, Dijon, France}}
\affil[4]{{\small Univ Angers, Nantes Universit\'e, CHU Angers, INSERM, CNRS, CRCI2NA, F-49000 Angers, France}}
\affil[5]{{\small Universit\'e Paris-Cit\'e, Hopital Saint Louis, INSERM U1131, F-75010 Paris, France}}

\date{}

\begin{document}

\maketitle

\noindent\textbf{Corresponding author}: Ana Fernández Baranda, ana.fernandez-baranda@polytechnique.edu \\

\noindent\textbf{Short title}: Modeling MPN development \\

\\

\noindent\textbf{Fundings}: Funded by the European Union (ERC, SINGER, n°101054787). Views and opinions expressed are however those of the author(s) only and do not necessarily reflect those of the European Union or the European Research Council Executive Agency. Neither the European Union nor the granting authority can be held responsible for them. \\
This work has also been supported by ITMO Cancer and by the Chair "Modélisation Mathématique et Biodiversité" of Veolia Environnement-Ecole Polytechnique-Museum National d'Histoire Naturelle-Fondation X.

\newpage

\section*{Abstract}
Mathematical modeling allows us to better understand myeloproliferative neoplasms (MPN), a group of blood cancers, emergence and development. We test different mathematical models on an initial cohort to determine the emergence and evolution times before diagnosis of JAK2V617F+ classical MPN (Polycythemia Vera (PV) and Essential Thrombocythemia (ET)). We consider the time before diagnosis as the sum of two independent periods: the time (from embryonic development) for the JAK2V617F mutation to occur, not disappear and enter proliferation, and a second time corresponding to the expansion of the clonal population until diagnosis. We prove that the rate of active mutation occurrence increases exponentially with age following the Gompertz model rather than being constant. We find that the first tumorous cell takes an average time of $63.1 \pm 13$ years to appear and start proliferation. On the other hand, the expansion time is constant: $8.8$ years once the mutation has emerged. These results are validated in an external cohort. Using this model, we analyze JAK2V617F ET versus PV, and obtain that the time of active mutation occurrence for PV takes approximately $1.5$ years more than for ET to develop, while the expansion time was similar. In conclusion, our age-dependent approach for the emergence and development of MPN demonstrates that the emergence of a JAKV617F mutation should be linked to an aging mechanism, and indicates a $8-9$ years period of time to develop a full MPN.

\newpage
\section{Introduction}
Myeloproliferative Neoplasms (MPN) are blood cancers issued from the transformation of hematopoietic stem cell and related to the acquisition of JAK2 signaling deregulation. The most frequent driver mutation in MPN is JAK2V617F, followed by JAK2 exon12, CALR and MPL \cite{Vainchenker}. MPN are categorized into three main diseases that present the JAK2V617F mutation in the majority of cases: Essential Thrombocythemia (ET), Polycythemia Vera (PV) and Primary Myelofibrosis (PMF) \cite{Tefferi}. ET is characterized by an increased platelet number, PV by erythroid compartment enlargement and PMF by bone marrow fibrosis. When the JAK2 mutation is responsible for the pathology development, the mutation is present in one allele for most cells for ET, but often undergoes an homologous recombination of JAK2V617F for PV resulting in the mutation being present in both alleles \cite{Scott}. In PMF, the JAK2 mutation is often coupled with secondary events such as TET2 mutation or ASXL1 \cite{Vainchenker}. We center our study on ET and PV, given the reduced number of reported cases for PMF. In order to develop one of these pathologies, two steps are needed: the acquisition of a mutation at stem cell level and the development of the cancerous population. The diagnosis of MPN just like Polycythemia Vera is performed when the total tumor mass (erythrocytes) represents at least $25\%$ of the erythroid hematopoietic mass \cite{Tefferi}. This suggests a total cancer mass of at least 750 grams in the blood and probably the same mass in the bone marrow. Based on the cell mass, we can speculate that a minimum of one million cell divisions are needed to create such a MPN. We also can speculate that the same order or tumor mass is present in all myeloid chronic myeloproliferative malignancies such as ET.

Mathematical modelling and statistical analysis based on patients data can help us better understand the dynamics of mutation emergence and expansion. This modelling could yield important information for prevention strategies and early diagnosis. Using real life registries of MPN and age of pathology detection in patients, we study the emergence and expansion of MPN for patients presenting the JAK2V617F mutation.

We define the time to diagnosis as composed of two independent elements: the time $T_1$ of active mutation emergence, namely the time for a first cell harboring the JAK2V617F mutation to emerge and enter the cell cycle, and the time $T_2$ between emergence and detection, namely the time for cancer population to grow until diagnosis. An important parameter in our model is the active mutation rate defined as the product of the mutation rate at stem cell level and the probability of each mutation to develop into a MPN. Based on a first cohort of diagnosis data, we prove here by a rigorous statistical analysis that this active mutation rate is age-dependent and increases exponentially following the Gompertz model (see \eqref{eq:gom}). We further prove that a constant MPN growing time is sufficient to properly explain the data. The statistical analysis of the model on a second cohort of diagnosis data validates the model and its robustness. The estimation of the parameters of the model allows to obtain information on characteristics of the diseases: an average time of $63.1 \pm 13$ years for the first tumorous cell to appear and start proliferation and a constant expansion time of $8.8$ years once the mutation has emerged. We also analyze JAK2V617F ET versus PV, and obtain that the time of active mutation occurrence for PV takes approximately $1.5$ years more than for ET to develop, while the expansion time is similar.

Mathematical modelling of MPN emergence and development based on patients data has already been introduced in different approaches. In \cite{Hermange2022} the authors use age of detection data to study the behaviour of the size of the mutant population. In other works, sequencing data is used to back-track mutations to define the timeline to driver mutation acquisition, (see for example \cite{Osorio2018}, \cite{VanEgeren2021}, \cite{Mitchell2022}). In all of these works around MPN emergence, an assumption of a constant mutation occurrence rate throughout the life of an individual is made. Such an assumption would imply an exponential distribution for the emergence time (see \autoref{subsec:cons}). However, the available data (\autoref{im:dataCO}) shows a growing number of diagnosed patients for older ages (see also \cite{Rohrbacher} and \cite{Moulard}) whose trend is not consistent with an exponential (see Figure \ref{im:M3}).

In \autoref{sec:framework} we describe the data used as well as the construction of the model. Section \ref{sec:model} is dedicated to introducing, selecting and validating our age-dependent model. Finally, in \autoref{sec:app} we apply the selected model to the data to obtain information on the mutation emergence and expansion of MPN and the difference in them for PV and ET.

\section{Framework} \label{sec:framework}
\subsection{Incidence and frequencies of MPN according to age} 
In this study, we work with data obtained from the cancer registry of the Côte d’Or in France from 1990 to 2020, which compiles the ages of detection for $264$ patients with different types of MPN (\autoref{table:data1} in Appendix). In this region, life expectancy (and then relative incidence of pathology) can be analyzed. We center our analysis to patients diagnosed with either PV or TE and presenting the JAK2V617F mutation. The mean detection age in this data set is of 71.9 years with a standard deviation of 13.8 years. To further validate the models, another set from FIMBANK was used, containing the ages of detection for ET and PV of 1111 individuals as seen in \autoref{table:data2} in Appendix. In this data set, we have a mean of 64.4 years with a standard deviation of 14.1 years.

\begin{figure}
    \centering
    \includegraphics[width = 0.8 \textwidth]{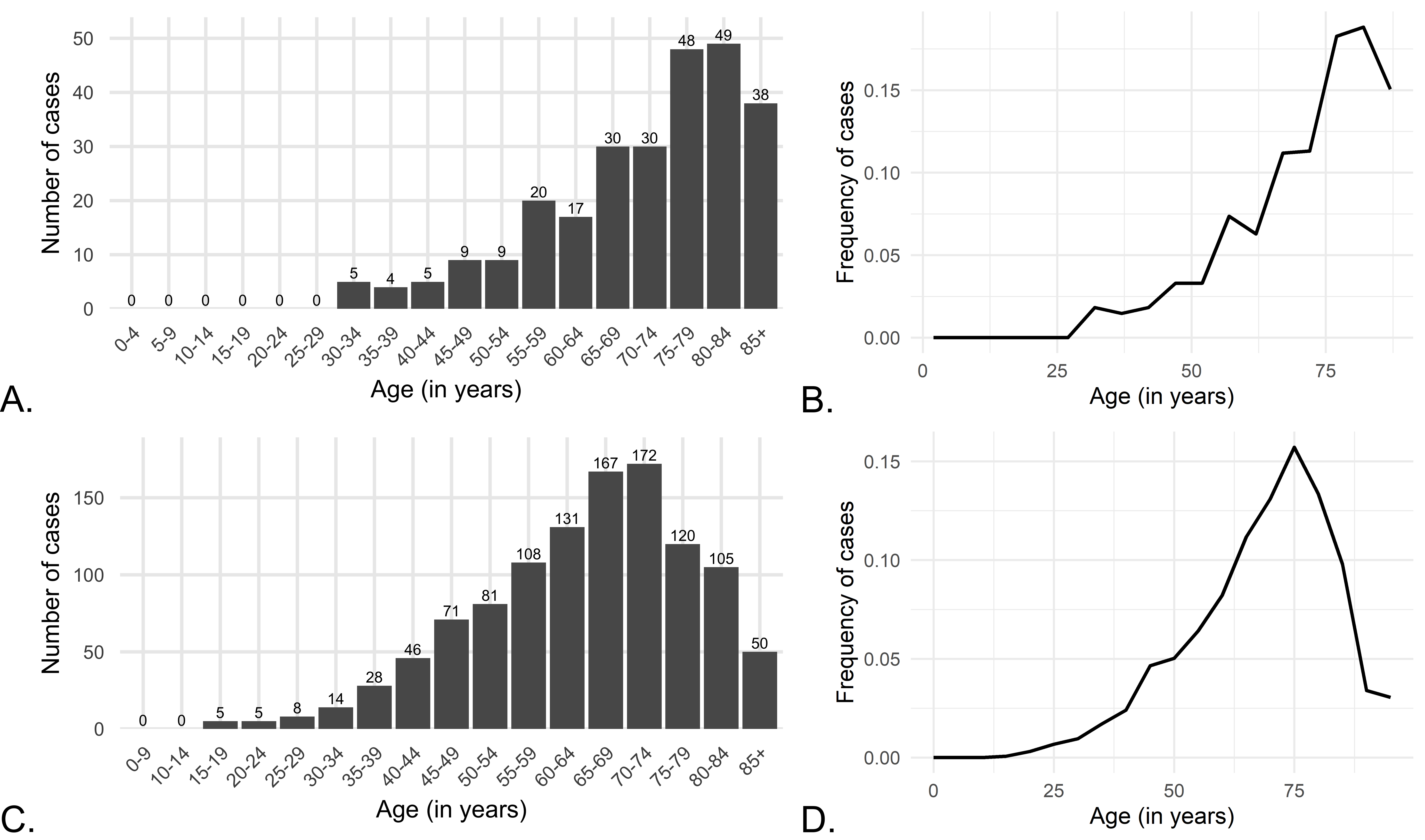}
    \caption{\textbf{A}: Incidence by age of JAK2V617F Classical MPN (ET and PV, excluding myelofibrosis) in the Cote d’Or Regional Registry Database. \textbf{B}: Incidence by age of JAK2V617F classical MPN (ET and PV, excluding myelofibrosis) in the Cote d’Or Regional Registry Database after adjustment excluding other causes of death before the MPN diagnosis. \textbf{C}: Incidence by age of JAK2V617F Classical MPN (ET and PV, excluding myelofibrosis) in the National BCB-FIMBANK Cohort. \textbf{D}: Incidence by age of JAK2V617F classical MPN (ET and PV, excluding myelofibrosis) in the National BCB-FIMBANK Cohort after adjustment excluding other causes of death before the MPN diagnosis.}
    \label{im:dataCO}
\end{figure}

The two sets of data used in our analysis are adjusted to take into account the distribution of the population into each age group and consider the proportion of individuals diagnosed within each group. This considers that some individuals could have died before being diagnosed and are thus not present in the data sets. Each frequency is then divided by the probability of survival up to that age group as follows. Let $T_D$ be the age of death and $T_M$ the age of detection. Then the frequencies given by the data of the probability of being diagnosed at age $t$ is given by
\begin{equation*}
    \mathbb{P}(T_M = t, T_D \geqslant t).
\end{equation*}
If we suppose independence between the two variables $T_D$ and $T_M$, this is equal to
\begin{equation*}
    \mathbb{P}(T_M = t) \mathbb{P}(T_D \geqslant t).
\end{equation*}
Using the death rates for each age group, one can obtain $\mathbb{P}(T_D \geqslant t)$ and then calculate
\begin{equation*}
    \mathbb{P}(T_M = t) = \frac{\mathbb{P}(T_M = t, T_D \geqslant t)}{\mathbb{P}(T_D \geqslant t)}.
\end{equation*}
We apply this adjustment based on the French and European survival data by age and sex of 2021 \cite{INED2021}. The resulting corrected data gives us the age distribution of JAK2V617F for patients who have not died of anything else before. This data indicates that the frequencies of JAK2V617F MPN development increases dramatically in older ages as shown in \autoref{im:dataCO}.

\subsection{Model construction and notation} \label{subsec:cons}
In order to perform a diagnosis of PV or ET, the total cancer mass needs to surpass a certain threshold of around $M = 10^{12}$ cells. We assume that diagnosis is performed when this size is reached. We consider that the age of detection, $T_M$, is the sum of two independent time periods, $T_1$ and $T_2$: the first one being the time from embryonic development for the mutation to occur, not disappear, and enter in proliferation (that we will name “mutation occurrence”) and a second time corresponding to the expansion time, once the first cancer cell begins to proliferate until the time of MPN diagnosis. The times $T_1$ and $T_2$ are taken as independent since the time taken for a cell to become proliferative should not be related to the time for its population to grow to a certain size. In order to consider that the mutation could have appeared at an embryonic level, as suggested in \cite{Mosca}, $T_M$ is considered to start 9 months before the birth of an individual.

To define the first period of time $T_1$, let $\tau$ be rate of mutation or activation of an existing mutation at stem cell level. Then, the time for a mutation to appear follows an exponential distribution with parameter $\tau$. We consider that each activated mutant cell has a probability $p$ to proliferate and then be detected. The number $(1 - p)$ corresponds to the frequency of non proliferating mutant cells (just like clonal hematopoiesis does for a long period of time) or of mutant cells whose lineages disappear before reaching detection size of MPN. To find the law of $T_1$ under these assumptions, let $L$ be a random variable of law $Geom(p)$ that corresponds to the number of mutation events before the first active cancer cell appears and let $X_1, X_2, ...$ be independent and identically distributed random variables of law $Exp(\tau)$ that define the time elapsed between two mutation attempts. Then, $T_1$ will simply be the sum of these times:
\begin{equation*}
    T_1 = \sum_{i = 1}^L X_i.
\end{equation*}
The law of this variable can be found using the generating function
\begin{align*}
    \mathbb{E}\left[e^{- \lambda T_1}\right] =& \mathbb{E}\left[\mathbb{E}\left[e^{- \lambda T_1} | L \right]\right] = \mathbb{E} \left[ \left( \frac{\tau}{\tau - \lambda} \right)^L \right] \\
    =& \sum_{k \leq 1} \left( \frac{\tau}{\tau - \lambda} \right)^k p (1 - p)^{k - 1} = \frac{\tau p}{\tau p - \lambda}.
\end{align*}
Then, $T_1$ is an exponential random variable with parameter $\delta = \tau p$, which we will refer to as active mutation rate. If we assume that $\delta$ depends on the age $t \geq 0$ of an individual, the density of $T_1$ is given by
\begin{equation} \label{eq:f1}
    f_1(t) = \delta(t) \exp\left(- \int_0^t \delta(s) \, ds \right).
\end{equation}

On the other hand, the time $T_2$ corresponds to the time taken for a population of cells to reach detection size of $10^{12}$ cells. Given the high number of divisions needed to reach this size ($10^6$ divisions), variability in division time is averaged and it makes sense to consider that the expansion time should be similar from one individual to the other. The law of $T_2$ then needs to be unimodal, centering around a certain value and positive as, for example, a lognormal law whose density for an age $t \geq 0$ is 
\begin{equation} \label{eq:T2log}
    f_2(t) = \frac{1}{t \sigma \sqrt{2 \pi}} \exp \left( - \frac{\left( \log(t) - \mu \right)}{2 \sigma^2} \right),
\end{equation}
with $\mu \in \mathbb{R}$, $\sigma > 0$. The simplest case is to consider a null variance with $T_2 = \alpha$ for $\alpha \geq 0$, where all individuals have the same expansion time, implying that environmental modifications do not have an effect on the growth of the tumor.

By independence of $T_1$ and $T_2$, the law of $T_M$ is computed as the convolution of the densities $f_1$ and $f_2$ of $T_1$ and $T_2$, respectively. That is, for $t \geq 0$, it is given by
\begin{equation} \label{eq:fM}
    f_M(t) = \int_0^t f_1(s) \, f_2(t-s) \, ds.
\end{equation}

The variables and parameters are summarized in the following table.

\begin{table}[h]
    \centering
    \begin{tabular}{ll}
        \toprule
        Variable & Description \\
        \hline
        $T_M$ & Time to detection. $T_M = T_1 + T_2$ \\
        $T_1$ & Mutation emergence time.\\
        $T_2$ & Expansion time.\\
        \toprule
        Parameter & Description \\
        \hline
        $\delta$ & Active mutation rate. $\delta = \tau p$\\
        $\tau$ & Mutation rate.\\
        $p$ & Probability of proliferation and detection of a mutant cell.\\
    \end{tabular}
    \caption{Variables and parameters of the model.}
    \label{table:params}
\end{table}

\section{Mathematical model and results} \label{sec:model} %and validation?
\subsection{Mathematical age-dependent model}

Data in \autoref{im:dataCO} shows a clear age structure, with an increasing number of cases as age increases. Then, we hypothesize an age-dependent active mutation occurrence rate that grows as a person gets older. Biological motivation behind considering an age-dependent mutation occurrence rate can be justified by at least three age-dependent modifications of the environment: i-JAK2V617F mutation appearance (from embryo to adulthood), ii- JAK2V617F activation (ie stem cell quiescence exit) or iii- the success probability of the expansion after activation. We consider the Gompertz model (see \cite{Lavielle2014}, pages 90-92) for the active mutation rate $\delta$, which is given by
\begin{align} \label{eq:gom}
    \delta_G(t) &= A \, \exp(k \, t),
\end{align}
where $A > 0$, $k > 0$, and $t$ is the age of an individual.
The Gompertz model has been widely used to model different biological age-dependent rates such as mortality \cite{Finch1990} or cancer death rate \cite{Riffenburgh2001}. In our case, it can be explained as follows: we assume that individuals have an initial mutation occurrence rate  $A$, and an age-related increase in mutation occurrence represented through a a parameter $k$. Then, for an age $t$, the active mutation occurrence takes the form $\delta_G(t)$ and the density of $T_1$ in \eqref{eq:f1} is expressed as
\begin{equation*}
    f_1(t) = A \, \exp \left( \frac{A}{k} \right) \, \exp(k \, t) \exp\left( - \frac{A}{k} \, \exp(k \, t) \right)
\end{equation*}

We consider two models for the expansion time. The first one being a fixed $T_2$ assuming that environmental modifications do not have an effect on the growth of the tumor. The second model takes $T_2$ as a lognormal random variable defined by \eqref{eq:T2log}. This setting can be explained by invasion process modifications (ie variation time between two divisions), environment modifications or randomness in the diagnostic process (some patients are diagnosed after vascular events (late detection) when others are diagnosed after a systematic blood cell analysis performed for another pathology (early detection)) or some subjects die (of other unrelated pathologies) before $T_2$ was finalized. Then, we have two age-dependent models characterized by
\begin{align}
    \text{Model A.1:}&
    \begin{cases} \label{eq:mod1}
        \delta_G(t) = A \, \exp(k \, t) \\
        T_2 = \alpha \text{ fixed},
    \end{cases}
    \\
    \text{Model A.2:}&
    \begin{cases} \label{eq:mod2}
        \delta_G(t) = A \, \exp(k \, t) \\
        T_2 \sim lognormal(\mu, \sigma^2).
    \end{cases}
\end{align}
The respective laws, $f_1^M$ and $f_2^M$, of the age of detection $T_M$ under each model come from \eqref{eq:fM} and are given for an age $t$ by
\begin{align*}
    f_1^M(t) &= A \, \exp \left( \frac{A}{k} \right) \, \exp(k(t-\alpha)) \exp\left( - \frac{A}{k} \, \exp(k(t - \alpha) \right) \\
    f_2^M(t) &= \int_0^t A \, \exp \left( \frac{A}{k} \right) \, \exp(k \, s) \exp\left( - \frac{A}{k} \, \exp(k \, s) \right) \frac{1}{t \sigma \sqrt{2 \pi}} \exp \left( - \frac{\left( \log(t - s) - \mu \right)}{2 \sigma^2} \right) ds.
\end{align*}

\subsection{Parameter estimation} \label{subsec:EM}
We estimate the parameters using the generalized Expectation-Maximization (EM) algorithm (see for example \cite{McLachlan2008}) as presented in \cite{Rai1993}. Let us recall the method. Let $T_M$ be the vector that corresponds to the observed data of detection ages and has distribution $f_{T_M}(t_M; \theta)$ for an age $t_M$, where $\theta$ is the vector of parameters ($\theta = (A, k, \alpha)$ for Model A.1 \eqref{eq:mod1} and $\theta = (A, k, \mu, \sigma^2)$ for Model A.2 \eqref{eq:mod2}). Consider the missing data $t_2$, corresponding to the values of $T_2$ with distribution $f_{T_2}(t_2; \theta)$. Then the complete data is $(t_M, t_2)$, and it is distributed according to $f_{T_M,T_2}(t_M, t_2 ; \theta)$. If the complete data was available, the log-likelihood could be calculated by
\begin{equation*}
    \log L_{T_M,T_2}(\theta) = \log f_{T_M,T_2}(t_m, t_2 ; \theta).
\end{equation*}
The goal is to find the values of the parameters that maximize this function. 
This algorithm is iterative and consists of two steps: the E-step and the M-step. Starting from an initial value $\theta_0$, then, on iteration $k + 1$, the E-step involves computing
\begin{equation*}
    Q(\theta; \theta_k) = \mathbb{E}_{\theta_k} ( \log L_{T_M,T_2}(\theta) | t_M).
\end{equation*}

Then, on the M-step choose $\theta_{k + 1}$ such that
\begin{equation*}
    Q(\theta_{k + 1}; \theta_k) \geqslant Q(\theta_k ; \theta_k).
\end{equation*}

For the M-step, as suggested in \cite{Rai1993}, we take $\theta_{k + 1}$ as one Newton-Raphson step from $\theta_k$ over the function $Q(\theta_{k + 1}; \theta_k)$, that is
\begin{equation*}
    \theta_{k + 1} = \theta_k + a_k \delta_k,
\end{equation*}
where
\begin{equation*}
    \delta_k = - \left. \left[\frac{\partial^2 (Q(\theta ; \theta_k))}{\partial \theta \, \partial \theta^T}\right]^{-1} \right|_{\theta = \theta_k} \left. \left[\frac{\partial (Q(\theta ; \theta_k))}{\partial \theta}\right] \right|_{\theta = \theta_k} 
\end{equation*}
and $0 < a_k \leqslant 1$. The choice of $a_k$ needs to be such that it stays in the parameters space and the likelihood is non-decreasing. To do this, take $a_k^0 = 1$ and divide by half its value until the parameters satisfy the desired conditions. Then, we apply backtracking line search: while $Q(\theta_k + a_k \delta_k; \theta_k) > Q(\theta_k; \theta_k) + 10^{-4} \, a_k \, \nabla Q(\theta_k; \theta_k)^T \, \delta_k$, set $a_k = 0.8 \, a_k$. Lastly, we take $a_k$ as the final value of these steps. 

\subsection{Model selection} \label{subsec:selec}
We want to compare and select a model amongst Model A.1 \eqref{eq:mod1} and Model A.2 \eqref{eq:mod2}. To do so, we estimate the parameters for both models using the EM-algorithm described in \autoref{subsec:EM} using the Côte d'Or cohort, and we proceed to perform a Chi-square goodness-of-fit test as follows. We consider the $k = 13$ bins corresponding to ages $0-30$, 5 year increments from age $30$ to $85$ and $85+$. For each bin $i$ ($i = 1, \dots, k$), let $O_i$ and $E_i$ be the number and the expected number of observations on bin $i$, respectively. Then, we construct the statistic as
\begin{equation*}
    \chi^2 = \sum_{i = 1}^k \frac{(O_i - E_i)^2}{E_i},
\end{equation*}
which follows approximately a Chi-square distribution with $k - c$ degrees of freedom, where $c$ is the number of estimated parameters ($c = 3$ for Model A.1 and $c = 4$ for Model A.2). 

\begin{figure}[h]
    \centering
    \includegraphics[width = 0.8 \textwidth]{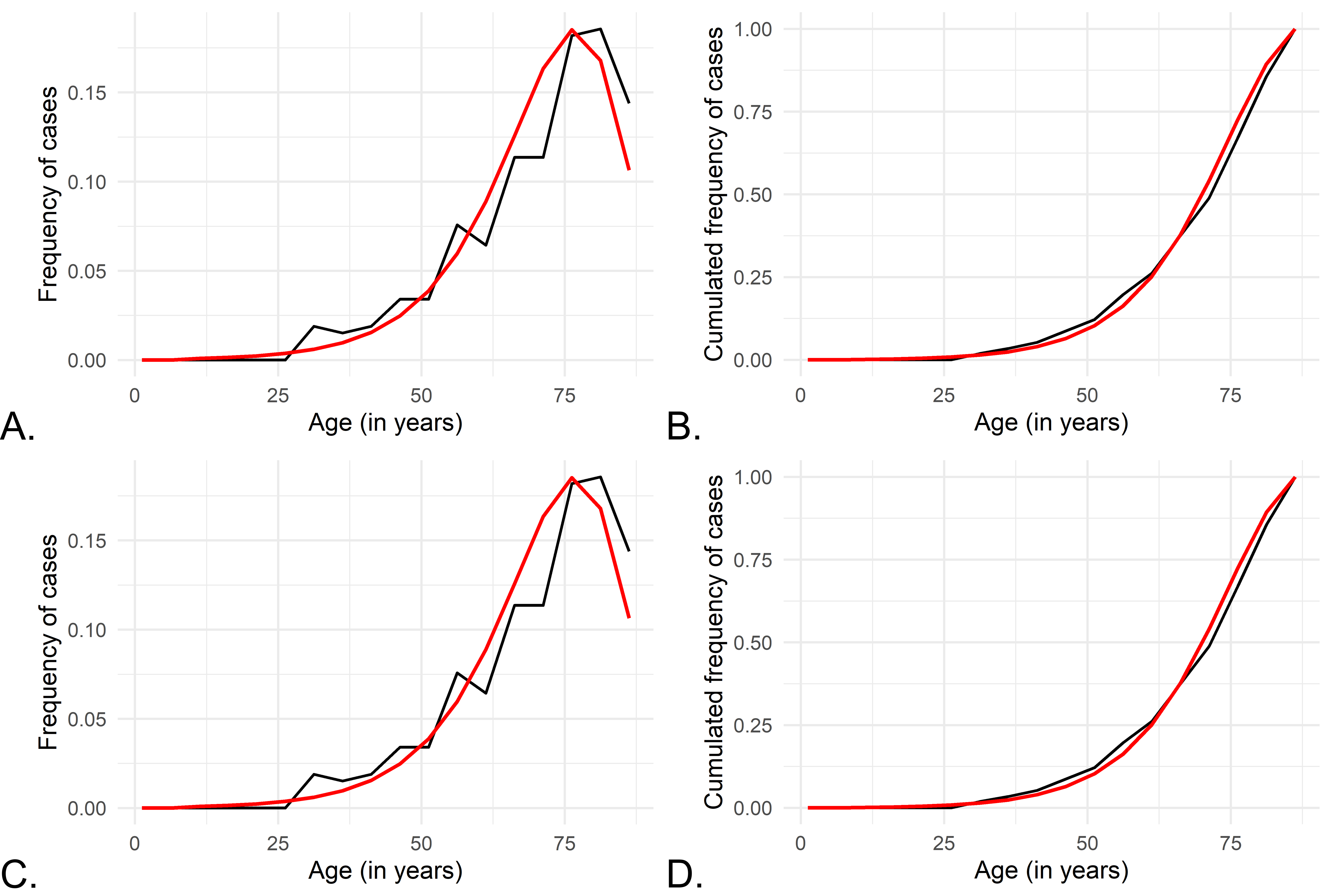}
    \caption{\textbf{A}: Frequency of cases for the data (in black) and probability density function of the estimation (in red) for the model with aging \eqref{eq:mod1}. \textbf{B}: Accumulated frequencies of cases for the data (black) and cumulative distribution of the estimation (in red) for the model with aging \eqref{eq:mod1}. \textbf{C}: Frequency of cases for the data (in black) and probability density function of the estimation (in red) for the model with aging and lognormal MPN growing time \eqref{eq:mod2} \textbf{D}: Accumulated frequencies of cases for the data (black) and cumulative distribution of the estimation (in red) for the model with aging and lognormal MPN growing time \eqref{eq:mod2}.}
    \label{im:M4}
\end{figure}

These tests resulted in not rejecting Model A.1 with a power of $0.952$ and not rejecting Model A.2 with a power of $0.85$. We refer to \autoref{im:M4} for visual comparison between the estimations for fixed $T_2$ in \textbf{A-B} and random $T_2$ in \textbf{C-D}. 

Since both models are not rejected, we use the Bayesian information criterion (BIC) presented in \cite{Schwarz1978} to select the most appropriate model amongst these two to explain the data. This criterion is defined by
\begin{equation*}
    \text{BIC} = c \log N - 2 \log \mathcal{L},
\end{equation*}
where $c$ is the number of parameters of the model, $N$ is the size of the data sample and $\mathcal{L}$ is the likelihood function. The BIC uses the likelihood function to measure how well a model can explain the data, while also penalizing the number of parameters in the model to avoid overfitting. A lower BIC is preferred. We obtain that the Model A.1 with constant MPN expansion time ($T_2$ as a fixed value) has a BIC of 2079 and Model A.2 with lognormal MPN expansion time has one of 2085.3. This indicates that the Model A.1 with fixed expansion time should be chosen. While incorporating variability on the expansion time $T_2$ gives a better fit to the data in terms of the likelihood, the gain is small and not compensated by the cost of adding additional parameter to the model. This would imply that, given our data, the Gompertz model with constant MPN growing time is the most appropriate to explain the data. This suggests that environmental modifications (or other sources of variability) do not have a huge impact on the growth phase of the tumor.

\subsection{Model validation}
\subsubsection{Validation on a second cohort}
\begin{figure}
    \centering
    \includegraphics[width = 0.8 \textwidth]{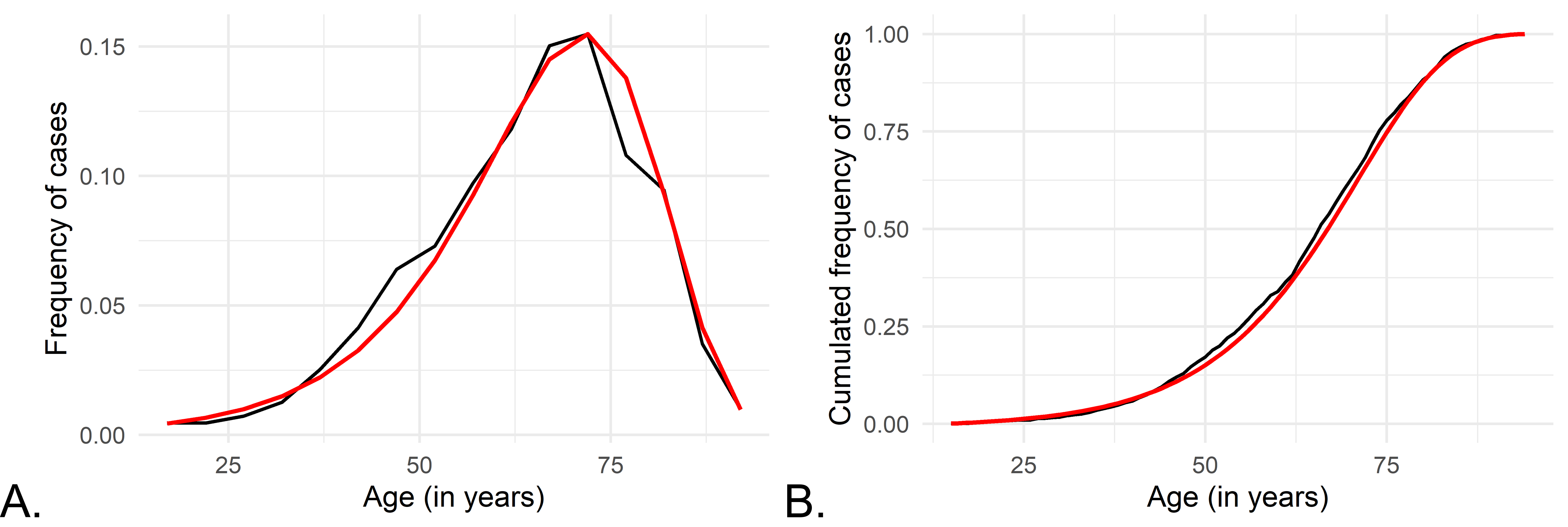}
    \caption{\textbf{A}: Incidence by age of JAK2V617F Classical MPN (ET and PV, excluding myelofibrosis) in the National BCB-FIMBANK Cohort. \textbf{B}: Incidence by age of JAK2V617F classical MPN (ET and PV, excluding myelofibrosis) in the National BCB-FIMBANK Cohort after adjustment excluding other causes of death before the MPN diagnosis. }
    \label{im:fim}
\end{figure}

The French group of MPN recently developed a national database on ET and PV (BCB-FIMBANK: French National Cancer Institute (INCa) BCB 2013 and 2022, CHU Angers) (Table 2). We considered that this database is representative of the general population given the data size ($1111$ JAK2V617F MPN patients) (see \autoref{im:dataCO} \textbf{C-D} for the histogram before and after death adjustment), even if it is not as exhaustive as the previous database that reported all the cases in the region. We used this second data set to validate the Gompertz model with fixed expansion time given by Model A.1 \eqref{eq:mod1}. We performed an estimation of the parameters through the same EM algorithm as previously described and then applied a Chi-Square goodness-of-fit test, which was not rejected with a power of $0.99$ (see \autoref{im:fim}). This model seems to properly explain MPN emergence and proliferation given our two data sets.

\subsubsection{Age-independent models and Weibull age-dependent model}
\begin{figure}
    \centering
    \includegraphics[width = 0.8 \textwidth]{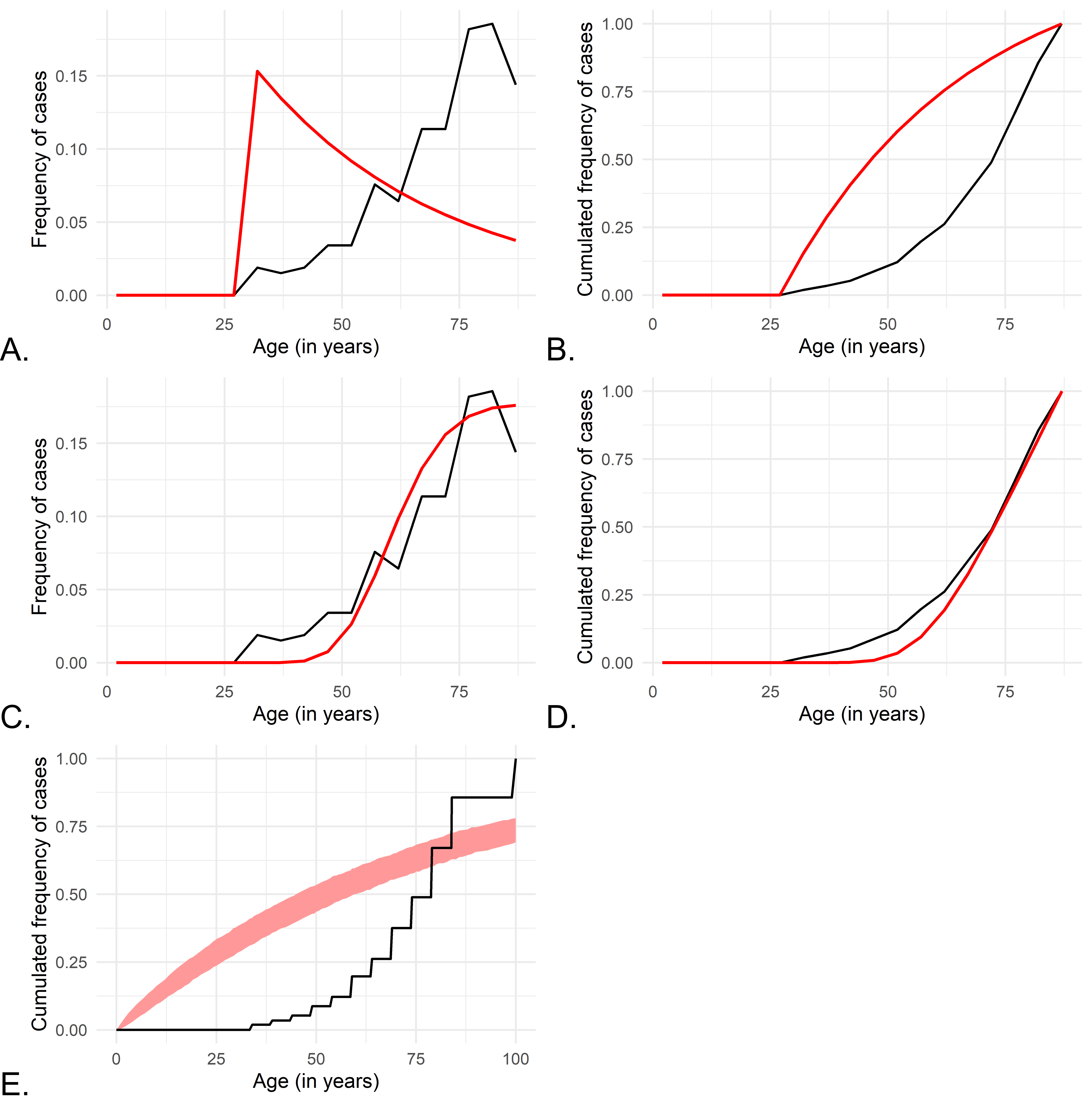}
    \caption{\textbf{A}: Frequency of cases for the data (in black) and probability density function of the estimation (in red) for the age-independent model (Model B.1 \eqref{eq:M31}). \textbf{B}: Accumulated frequencies of cases for the data from the (black) and cumulative distribution of the estimation (in red) for the age-independent model (Model B.1 \eqref{eq:M31}). \textbf{C}: Frequency of cases for the data (in black) and probability density function of the estimation (in red) for the age-independent model with lognormal MPN growing time (Model B.2 \eqref{eq:M32}). \textbf{D}: Accumulated frequencies of cases for the data (in black) and cumulative density of the estimation (in red) for the age-independent model with lognormal MPN growing time (Model B.2 \eqref{eq:M32}). \textbf{E}: Accumulated frequencies of cases for the data (in black) and $90\%$ confidence interval of the estimation (in red) for the age-independent model with individual variability (Model B.3 \eqref{eq:M33})}
    \label{im:M3}
\end{figure}

The proposed age-dependent model was selected among others given its performance with the data. Indeed, when testing different models for the emergence and expansion time ($T_1$ and $T_2$), the models fail to fit the data. We study three models with a constant active mutation rate $\delta$, which implies $T_1 \sim Exp(\delta)$ (see \autoref{subsec:cons}). This assumption of a constant mutation rate has been taken in the majority of genetic analysis backtracking reported to date \cite{Osorio2018}, \cite{VanEgeren2021}, \cite{Mitchell2022}. The first and most simple model considers a fixed expansion time $T_2$. A second model considers that external factors may influence the expansion time, $T_2$ and is hence taken as a lognormal random variable as defined in \eqref{eq:T2log}. The third model assumes a fixed expansion time, but allows the time of mutation occurrence to vary from one patient to another or the mutation to develop at different times from one patient to another (quiescence/cycling). This takes into consideration that the biological hypothesis of stress (or environment) influences cancer cell emergence. This model considers $T_1$ as an exponential random variable, but its rate is different from one patient to another. Then, for $1 \leq i \leq N$, where $N$ is the number of patients, we considered that each age of emergence $T_1^i$ is drawn from an exponential distribution with parameter $\delta_i$. The $\delta_i$'s are considered as an independent sample of a $lognormal (m, s^2)$.
The three age-independent models can be summarized as
\begin{align}
    \text{Model B.1}&: 
    \begin{cases} \label{eq:M31}
        T_1 \sim Exp(\delta) \\
        T_2 = \alpha \text{ fixed},
    \end{cases}
    \\
    \text{Model B.2}&: 
    \begin{cases} \label{eq:M32}
        T_1 \sim Exp(\delta) \\
        T_2 \sim lognormal(\mu, \sigma^2),
    \end{cases}
    \\
    \text{Model B.3}&: 
    \begin{cases} \label{eq:M33}
        T_1^i \text{ sampled from } Exp(\delta_i)\\
        \delta_i \text{ sampled from } lognormal (m, s^2), \ i = 1, \dots, N \\
        T_2 = \alpha \text{ fixed}.
    \end{cases}
\end{align}
We estimate the parameters of Model B.1 and the ones of Model B.2 through a least-squared estimation which minimizes the sum of the square difference between the data and the model for each age. The parameters of Model B.3 are estimated using the software Monolix \cite{Lavielle2014}. This software is well adapted for this mixed effect modeling as it allows to easily estimate the parameters when considering the survival analysis approach. The estimation is based of the SAEM algorithm and gives robust and global convergence. As illustrated in Figure \ref{im:M3} \textbf{A}, \textbf{B} and \textbf{E}, the approximation for Model B.1 and Model B.3 presents such a big discrepancy with the data that it is sufficient to reject this model. While Model B.2 seems closer to the data, we perform a Chi-square goodness-of-fit test which results in rejecting this model at a $99\%$ confidence level with a p-value smaller than $10^{-17}$. The rejection of all three of these models imply that even when including individual or environmental variations, models with constant active mutation occurrence $\delta$ cannot explain the data.

We can then think of other age-dependent models, such as a Weibull model for the active mutation rate (see \cite{Lavielle2014}, pages 90-92). For an age $t$, the Weibull mutation occurrence rate function is given by
\begin{align*}
    \delta_W (t) &= \frac{\gamma}{\lambda} \left( \frac{t}{\lambda} \right)^{\gamma - 1},
\end{align*}
where $\gamma > 0$, $\lambda > 0$. We consider the broader model with $T_2$ as a lognormal random variable with density function \eqref{eq:T2log}. The addition of variability to the expansion time $T_2$ should give the same fit or better than a fixed value. The parameters are estimated using the EM-algorithm in \autoref{subsec:EM}. Then, we perform a Chi-square goodness-of fit test and find that the Weibull model is rejected with a p-value smaller than $10^{-17}$. 
\\ \ \\

Hence, we obtain a good fit of Model A.1 with Gompertz age-dependent active mutation rate and fixed expansion time in both data sets. Coupled with the rejection of all three age-independent mutation rate models and the Weibull age-dependent model, these results suggest that not only the mutation rate cannot be constant through life, but increases exponentially with age following the Gompertz model. Given these positive results to fit the data, we apply Model A.1 \eqref{eq:mod1} to our data to obtain information on the emergence and expansion of the disease and to compare the differences in mutation acquisition and expansion between the two pathologies PV an ET.

\section{Applications of the model} \label{sec:app} %Results?
\subsection{Mutation emergence and expansion times}

By estimating the parameters of Model A.1 \eqref{eq:mod1} to the Côte d'Or data set, using the EM-algorithm (\autoref{subsec:EM}), we obtain the following values
\begin{equation*}
    A = 0.000124, \quad k = 0.096577, \quad \alpha = 8.76.
\end{equation*}
This implies that the expected occurrence time $T_1$ is of $63$ years with a standard deviation of $13$ years and the pathology development when the first JAK2V617F stem cell is present, fixed and ready to proliferate is $8.8$ years ($T_2$), with a mean detection age of $76$ years. 

\subsection{Differential analysis of Essential Thrombocythemia and Polycythemia Vera}
We hypothesize that ET and PV have a different kinetic because of the double hit in PV compared to single hit in ET: homologous recombination or development of secondary mutations that induces the polycythemia instead of the thrombocytosis only. Given the high number of reported MPN in the FIMBANK cohort, we use this data set to analyze the difference between ET and PV patients. 

\begin{figure}
    \centering
    \includegraphics[width = 0.4 \textwidth]{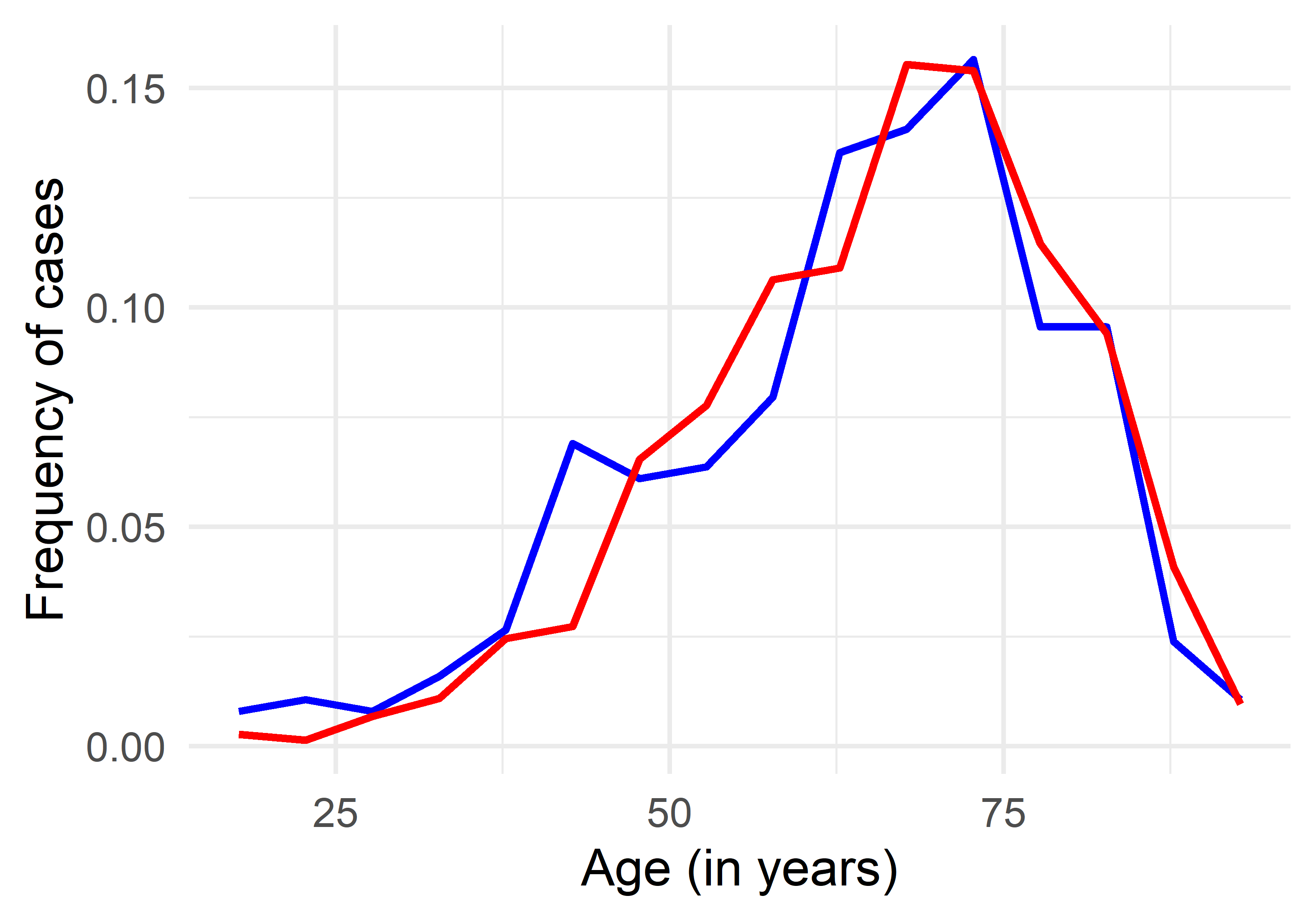}
    \caption{Frequencies of JAK2V617F ET (red line) and JAK2V617F PV (blue line) by age at diagnosis in the National BCB-FIMBANK Cohort}
    \label{im:pvte}
\end{figure}

There is not a clear difference between the two diseases data (see \autoref{im:pvte}). Yet, when looking at their means and standard deviations (see \autoref{table:pvet}),  PV presents a bigger mean by 2 years than ET. The standard deviation of ET is bigger by 1.6 years, meaning that the ages of detection are spread across a larger range. 
We start by testing if ET and PV corrected data come from the same distribution through a Chi-square test, which rejects this hypothesis with a p-value of $10^{-11}$. Hence, we confirm that the diagnostic age distributions are different for ET and PV (at a 99\% confidence level). 

We proceed to estimate the parameters of each pathology under Model A.1 \eqref{eq:mod1} and find the following values for the parameters, the mean ($m_1$) and standard deviation ($s_1$) of $T_1$ for each disease
\begin{gather*}
    A^{PV} = 0.00037, \quad k^{PV} = 0.08575, \quad T_2^{PV} = 8.8, \quad m_1^{PV} = 56.9, \quad s_1^{PV} = 14.2 \\
    A^{ET} = 0.00056, \quad k^{ET} = 0.07935, \quad T_2^{ET} = 8.3, \quad m_1^{ET} = 55.5, \quad s_1^{ET} = 15.1.
\end{gather*}
The estimation under our model indicates that the average mutation occurrence time ($T_1$) is approximately 1.5 years higher for PV which can be explained by the time taken for the mitotic recombination to happen.

\section{Discussion}
Our present work provides mathematical modeling of JAK2V617F MPN emergence and expansion based on real life incidence and age at diagnosis, which highlights the role of an exponential age dependence increase of mutation occurrence in MPN. Based on the regional exhaustive registry of Côte d’Or (France), and lastly validated in a MPN national registry (BCB-FIMBANK registry), we model the occurrence and expansion times of JAK2V617F MPN. Most of the previous works performed to decipher the time of JAK2V617F “first” cell appearance, are based on genetic hierarchy and back-tracking performed on ancient blood samples or bone marrow samples (see \cite{Osorio2018}, \cite{VanEgeren2021}, \cite{Mitchell2022}). These genetic approaches rely upon premises of a rate (a frequency) of mutation occurrence as yet unknown in MPN. It necessitates old blood or bone marrow samples for genetic analysis (to recapitulate the genetic tree evolution of the initial cancer cell) which is obviously very rare and probably biased \cite{Hirsch2016}. Genetic analysis inference performed in such approaches are based on mutation appearance but not necessary proliferative/oncogenic. Indeed, it is now well documented that such JAK2V617F mutations could stay for long time without development of pathologies, as seen in clonal hematopoiesis of indeterminate potential (CHIP). Moreover, these genetic back-tracking approaches were built on stable mutation frequencies that overestimate the time of emergence if mutation rate during life is not constant. Our approach is based on registry analysis, differing from classical genetic back-tracking of mutation emergence and we show that not only the mutation rate cannot be constant, but needs to grow exponentially with age, following the well-known Gompertz model.
We assume that two periods of time are necessary to develop and diagnose MPN: a first time period ($T_1$) corresponding to the mutation acquisition (genetic event from the conception of the embryo) but also its activation (ie ensuring to develop). This first period of time can correspond biologically to a cell cycle entry of a long-lasting quiescent cell for example, but also includes the possibility of a MPN cancer cell to disappear due to the cancer cell extinction probability. The second time period ($T_2$) corresponds to the growth and development of such cancer stem cell until diagnosis. This type of modeling has been applied to invasion-fixation of mutants in eco-evolutionary models (see \cite{Champagnat2006}, \cite{Billiard2017}) and to the infectious diseases propagation (see \cite{Barbour2015}, \cite{Barbour2013}) but less frequently to cancer emergence and development.
Most reported modeling assumes that mutation and invasion rate are aging independent (ie a fixed rate of mutation and activation and a fixed probability of emergence of the pathology for a given patient), which would imply that $T_1$ follows an exponential law. However, models under this hypothesis show a peak of cases in younger ages, which is not coherent with the data, where most cases are present for older ages. We then speculate that myeloproliferative neoplasm emergence could need an age-dependent modeling for $T_1$. Age dependence does not necessarily mean that the JAK2V617F mutation acquisition is higher in aged stem cells than in young ones; it also could be interpreted as a mutation appearance during childhood or even before birth, but an activation due to external stimulus long time after. The second time (expansion time) is easier to characterize. A model considering the expansion time  as a fixed value properly fits the data and is chosen over one as a random variable. This suggests that external changes (stress, food, toxic exposure, ...) could not significantly modify the behavior of MPN development and that the proliferation time from one “active” cancer cell to the total mass of cancer at diagnosis is relatively fixed. This could mean that the main determining factor is the proliferative speed related to the mutation acquisition. We then can speculate about the possibility (and risks and benefits) to treat subjects with JAK2V617F clonal hematopoiesis to delay the emergence of the MPN or prevent the classical thrombotic complications related to these pathologies. Our mathematical and statistical study allows us to assess the time of emergence and development of MPN. We find that the mean time for cancer cell to emerge is 63 years and time from tumor emergence to diagnosis is approximately 8.8 years in accordance with previous hypothesis, (see \cite{Radivoyevitch2012} and \cite{McKerrell2017}). Based on the Variant Allele Frequency of JAK2V617F in the “FIMBANK” cohort, we observe that the mean JAK2V617F VAF in MPN is around 30\% (19.9\% in ET and 46.1\% in PV respectively) and an increased ratio of 1.4\% per year (as reported previously in untreated \cite{Stein2010} or Hydroxyurea only treated patients \cite{Kiladjian2022}. This should mean a 10-year period of time from undetectable VAF to the 30\% allele ratio observed at diagnosis in our MPN cohort. This is very close (and a little bit higher) to the 8.8 years of tumor expansion time to MPN diagnosis that we calculate in our model. Furthermore, our modeling fits well with the JAK2V617F CHIP that has been described to take approximatively a decade to transform to MPN \cite{Nielsen2011}, \cite{Nielsen2013}, \cite{vanZeventer2023}. In conclusion, we demonstrate using mathematical modelling that the emergence and fixation of JAK2V617F mutation is strongly linked to aging mechanisms and that the expansion time of MPN can be considered as non random. Our numerical results show a mean of 63.1 years for the emergence and fixation of the mutation, and an expansion time of approximately 8.8 years. Altogether, these results highlight the interest to test all 50-60 years-old inhabitants for JAK2V617F CHIP and the place of preventive therapies for such subjects.

\bibliographystyle{plainnat}
\bibliography{bibliography}

%\printbibliography

\newpage 

\appendix

\section*{Tables}

\begin{table}[h]
    \centering
    \begin{tabular}{lcccc}
        \toprule
        Age & PMF & PV JAK2+ & ET JAK2+ & Total JAK2+ cases\\
        \hline
                0 to 4 & 0 & 0 & 0 & 0 \\
        5 to 9 & 0 & 0 & 0 & 0 \\
        10 to 14 & 0 & 0 & 0 & 0 \\
        15 to 19 & 0 & 0 & 0 & 0 \\
        20 to 24 & 0 & 0 & 0 & 0 \\
        25 to 29 & 0 & 0 & 0 & 0 \\
        30 to 34 & 0 & 1 & 4 & 5 \\
        35 to 39 & 0 & 0 & 4 & 4 \\
        40 to 44 & 0 & 1 & 4 & 5 \\
        45 to 49 & 0 & 5 & 4 & 9 \\
        50 to 54 & 0 & 2 & 7 & 9 \\
        55 to 59 & 1 & 8 & 11 & 20 \\
        60 to 64 & 1 & 10 & 6 & 17 \\
        65 to 69 & 2 & 13 & 15 & 30 \\
        70 to 74 & 4 & 12 & 14 & 30 \\
        75 to 79 & 5 & 14 & 29 & 48 \\
        80 to 85 & 4 & 15 & 30 & 49 \\
        More than 85 & 4 & 11 & 23 & 38 \\
        \hline
        Total number & 21 & 92 & 151 & 264
    \end{tabular}
    \caption{First data set (Côte d'Or).}
    \label{table:data1}
\end{table}

\begin{table}[h]
    \centering
    \begin{tabular}{lccc}
        \toprule
        Age & PV JAK2+ & ET JAK2+ & Total JAK2+ cases \\ \hline
        0 to 4 & 0 & 0 & 0 \\
        5 to 9 & 0 & 0 & 0 \\
        10 to 14 & 0 & 0 & 0 \\
        15 to 19 & 2 & 3 & 5 \\
        20 to 24 & 1 & 4 & 5 \\
        25 to 29 & 5 & 3 & 8 \\
        30 to 34 & 8 & 6 & 14 \\
        35 to 39 & 18 & 10 & 28 \\
        40 to 44 & 20 & 26 & 46 \\
        45 to 49 & 48 & 23 & 71 \\
        50 to 54 & 57 & 24 & 81 \\
        55 to 59 & 78 & 30 & 108 \\
        60 to 64 & 80 & 51 & 131 \\
        65 to 69 & 114 & 53 & 167 \\
        70 to 74 & 113 & 59 & 172 \\
        75 to 79 & 84 & 36 & 120 \\
        80 to 84 & 69 & 36 & 105 \\
        85 to 90 & 30 & 9 & 39 \\
        90 to 94 & 7 & 4 & 11 \\
        \hline
        Total number & 734 & 377 & 1111
    \end{tabular}
    \caption{Second data set (FIMBANK).}
    \label{table:data2}
\end{table}

\begin{table}[h] 
    \begin{tabular}{llrrrrr} 
         &  & \begin{tabular}[c]{@{}l@{}}Number \\ of cases\end{tabular} & \begin{tabular}[c]{@{}l@{}}Mean age \\ at diagnosis\end{tabular} & \begin{tabular}[c]{@{}l@{}}Standard \\ deviation of age\end{tabular} & \begin{tabular}[c]{@{}l@{}}VAF JAK2 \\ at diagnosis\end{tabular} & \begin{tabular}[c]{@{}l@{}}Standard \\ deviation of VAF\end{tabular} \\
        \hline
        \multicolumn{1}{c}{\multirow{2}{*}{PV}} & Female & 314 & 66.7 & 13.7 & 45.3 & 22.9 \\
        \multicolumn{1}{c}{} & Male & 421 & 64.0 & 13.3 & 46.7 & 30.3 \\
        \hline
        \multirow{2}{*}{ET} & Female & 229 & 63.0 & 15.9 & 19.2 & 5.7 \\
         & Male & 147 & 63.2 & 13.8 & 20.9 & 15.8
    \end{tabular}
    \caption{Information concerning the pathologies PV versus ET by sex.} \label{table:pvet}
\end{table}
\end{document}